\documentclass{JHEP3}
\usepackage{url}
\usepackage{amsmath}
\usepackage{amsfonts}
\usepackage{amssymb}
\usepackage{graphicx}    
\usepackage{psfrag}
\usepackage{subfigure}

\newcommand{\dd}{\mathrm{d}}
\newcommand{\ee}[1]{\mathrm{e}^{#1}}
\newcommand{\beq}{\begin{equation}}
\newcommand{\eeq}{\end{equation}}

\author{Jaume L\'opez Carballo\\
Departament ECM, Facultat de F\'{\i}sica, Universitat de Barcelona,  Spain.\\
E-mail: \email{jlopez@ecm.ub.es}}

\author{Jorge G. Russo\\
Instituci\'o Catalana de Recerca i Estudis Avan\c cats (ICREA)\\
Departament ECM, Facultat de F\'{\i}sica, Universitat de Barcelona,  Spain.\\
E-mail: \email{jrusso@ecm.ub.es}}


\abstract{
We describe  new bubble decays in pure $D+1$ dimensional Einstein theory 
with two compact directions. The instanton solution is constructed by
analytic continuation of the Kaluza-Klein electrically charged
black hole solution.
We show that the instanton describes the decay of a Kaluza-Klein
vacuum $\mathbb M^{D-1} \times T^2$ with a non-vanishing torus tilt parameter. 
The decay is produced by the creation of a bubble of nothing 
which expands with time.
We compute the instanton action, which shows that this Kaluza-Klein vacuum
becomes more stable as the torus tilt parameter is increased.
As an application, we consider the decay of M-theory torus compactifications leading
to type 0A/0B string theories. 
}

\title{New Bubble Decays In Kaluza-Klein Theories}

\preprint{UB-ECM-PF-05/13}

\begin{document}

\section{Introduction}



Long ago Witten~\cite{Witten:1981gj} realized that the $\mathbb M^4
\times S^1$ vacuum of the original Kaluza-Klein theory is 
unstable via a semiclassical decay
~\cite{Coleman:1977py,Coleman:1980aw}
into a \emph{bubble of nothing},
where a hole forms in space, which rapidly expands to infinity,
reaching  the speed of light after a short time
(we use the notation $\mathbb M^n$ for Minkowski space in $n$ dimensions). 

To construct the bubble solution, Witten starts with  
the Euclidean version of the $D=5$ Schwarzschild black hole.
The Euclidean time, $\tau = i t$, now plays the role of the compact Kaluza-Klein coordinate.
The resulting metric represents an instanton solution that interpolates 
between the false (unstable) vacuum $\mathbb M^4 \times S^1$ at $r \gg
r_h$, and the true vacuum solution, which  is obtained by a
 second Wick rotation, leading to the bubble solution
\begin{equation}\label{eq:sch2wicks}
\dd s^2 = \dfrac{\dd r^2}{1-\Big(r_h/r\Big)^{\!2}} + \left(1-\Big(r_h/r\Big)^{\!2}\right) \dd \tau^2 - r^2 \dd\psi^2 + r^2 \cosh^2 \psi \ \dd\Omega_2^2 \ .
\end{equation}

A similar analytic continuation procedure can be applied to construct
a different instanton solution starting with the $D=5$ Myers-Perry Kerr
solution. By suitable Wick rotations, one obtains an Euclidean
solution of finite action,
\begin{equation}\label{eq:DowkerMetric}
\begin{split}
\dd s^2 = & \dd \tau^2 + \sin^2\theta ( r^2 - a^2 ) \dd\varphi^2 - \frac{\mu}{\rho} \Big( \dd\tau + a \sin^2 \theta \dd\varphi \Big)^{\!2} \\
& + \frac{\rho^2 \dd r^2}{r^2 - a^2 - \mu} + \rho^2 \dd\theta^2 + r^2 \cos^2 \theta \dd\chi^2 \ ,
\end{split}
\end{equation}
where $\rho^2 = r^2 - a^2 \cos^2\theta$.
A natural question is then what kind of decay this instanton describes.
This question was answered by
Dowker et al~\cite{Dowker:1995}, who showed that the instanton
solution
has the right asymptotics to describe the decay of Kaluza-Klein vacuum
in the presence of a magnetic field,  the  Kaluza-Klein Melvin background,
representing a static, cylindrically symmetric magnetic flux tube. 
Indeed, at infinity one has a flat metric, which in cylindrical coordinates
reads
\begin{equation}
	\dd s^2 = - \dd t^2 + \dd z^2 + \rho^2 \dd\varphi^2 + \dd\tau^2 \ .
\end{equation}
Demanding that the metric  \eqref{eq:DowkerMetric} is free of conical 
singularities at the horizon $r_h^2 = \mu + a^2$ leads to the 
identifications 
%
\begin{equation}\label{eq:identDowker}
(\varphi,\,\tau) = \Big(\varphi+2\pi n R_\tau B + 2\pi m, \, \tau + 2\pi n R_\tau \Big)
\end{equation}
with $n$, $m \in \mathbb Z$ and $B=a/\mu $,  which define the Melvin magnetic
flux tube. There are two instanton solutions \eqref{eq:DowkerMetric}
that approach the same Melvin
magnetic configuration, differing by a shift in the magnetic parameter
$B$. They describe two decay modes of this space.
One is the instability  via nucleation of a pair of monopoles
($B=a/\mu +1/R_\tau$), by a
magnetic dual of the Schwinger effect of pair creation in an electric field. 
The second decay mode ($B=a/\mu$) is 
the formation of a bubble of nothing.\footnote{In string theory,
there is in addition a tachyon instability originating from a winding mode 
which appears above some critical magnetic field parameter \cite{RT}.}

Having shown that $D=5$ Schwarzschild solution and
the $D=5$ Kerr solution describe instanton decays of
Kaluza-Klein vacua, a natural question is whether
a similar interpretation holds for the instanton constructed from the
Kaluza-Klein electrically charged
solution, i.e. a black hole with $U(1)$ charge associated
with a Kaluza-Klein gauge field.
The advantage of having a charge of Kaluza-Klein origin is that
in the higher dimensional theory the instanton is a solution of pure
Einstein theory.
Here we will show that this
instanton describes
 the semiclassical decay of a Kaluza-Klein space
$\mathbb M^4 \times T^2$ with a non-zero torus tilt, 
via the formation of a bubble of nothing.
The torus tilt parameter arises 
as a result of identifications which are necessary 
to render the space free of conical singularities.



\section{Bubbles decays from Kaluza-Klein charged solutions}\label{secc:solucion}

\subsection{$D=5$ black holes with Kaluza-Klein charge}

We consider pure Einstein theory in six dimensions.
We are interested in semiclassical decays of the vacuum
 $\mathbb M^4 \times T^2$
(in the appendix, we consider the case  
$\mathbb M^{D-1} \times T^2$).
One vacuum decay  is represented by the Witten bubble of nothing generalized
by including an extra $S^1$ dimension. 
This is constructed by adding to
the five dimensional Schwarzschild black hole metric an extra (isometric)
direction $y$, where $y$ is a periodic coordinate, \hbox{$y=y+2\pi R_y$}.

To construct more general decays,
our starting point is the electrically charged solution given
in~\cite{Gibbons:1985ac,Gibbons:1987ps}. 
The five-dimensional
solution is obtained as the Kaluza-Klein reduction of a
six-dimensional metric which is a solution of the 
Einstein equations in six dimensions, i.e. it satisfies 
$R_{\mu\nu}(g_{6})=0$. \footnote{A discussion (which has 
no overlap with the present treatment) of
bubble solutions related to four-dimensional Kaluza-Klein
charged black holes is in \cite{emparan}.
Discussions of other Kaluza-Klein bubble solutions can be found 
in \cite{Elvang,Reall}.
}

The six-dimensional metric is given by
\begin{equation}\label{eq:met6}
\begin{aligned}
\dd s^2_6 = & 
- \dfrac{1}{1+\alpha/r^2} \left[ \left( 1 - \big( r_h / r \big)^2 \right) - \dfrac{Q^2}{8\pi^4 r^4} \right] \dd t^2
+ \dfrac{\dd r^2}{1 - \big( r_h / r \big)^2}
\\ &
+ r^2 \dd\Omega_3^2 
+ \Big(1+\alpha/r^2\Big) \dd y^2
+ \dfrac{Q}{\sqrt 2 \pi^2 r^2} \dd y \dd t \ .
\end{aligned}
\end{equation}
The coordinate $y$ is periodic with $y=y+2\pi R_y$.
The dimensional reduction to five dimensions in $y$ is obtained
by writing the  metric in the form 
\begin{equation}\label{eq:cast6}
\dd s_6^2 = \ee{2\phi/3} \dd s_5^2 + \ee{-2\phi} \Big( \dd y + A_\mu \dd x^\mu \Big)^{\!2} \ .
\end{equation}
One finds that the five dimensional metric describe a black hole with mass $M$ and charge $Q$. Its line element is
\begin{subequations}
\begin{equation}\label{eq:met5}
\dd s_5^2 = - \dfrac{1-\big( r_h / r \big)^{\!2}}{\Big(1+\alpha/r^2\Big)^{\!2/3}} \dd t^2 + \dfrac{\Big(1+\alpha/r^2\Big)^{\!1/3}}{1-\big( r_h / r \big)^{\!2}} \dd r^2 + \Big(1+\alpha/r^2\Big)^{\!1/3} r^2 \dd\Omega^2_3 \ .
\end{equation}
The gauge field is given by $A_\mu = (A_0, 0, 0, 0, 0)$, where
\begin{equation}
A_0 = \dfrac{Q}{2\sqrt2\pi^2} \dfrac{1}{r^2 \Big(1+\alpha/r^2\Big)} \ .
\end{equation}
%
Finally, the scalar field is
\begin{equation}
\phi = - \frac12 \ln\Big(1+\alpha/r^2\Big) \ .
\end{equation}
\end{subequations}
The solution is fully specified by two parameters, namely the mass $M$
and the charge $Q$. Introducing a parameter $k$
\begin{equation}
k = 1 - \sqrt{1 + \frac32 \left( \frac{Q}{\kappa^2 M} \right)^{\!\!2}} \ ,
\end{equation}
the different parameters can then be written in terms of 
$M$ and $k$ by means of the following relations
\begin{subequations}
\begin{equation}
\begin{aligned}\label{QK}
Q^2 & =\frac23 (\kappa^2 M)^2 k \Big(k-2\Big) \ ,\nonumber\\
r_h^2 & = \dfrac{\kappa^2 M}{3\pi^2} \left( 1 + k \right) \ ,
\nonumber \\
\alpha & = - \dfrac{\kappa^2 M}{2\pi^2} \, k \ .
\end{aligned}
\end{equation}
\end{subequations}
The solution represents a black hole with regular horizon in the region
\begin{equation}
0\le \left( \frac{Q}{\kappa^2 M} \right)^{\!\!2} \le 2 \ .
\end{equation}
This corresponds to the range $-1\leq k\leq 0$.
For larger values of $Q/M$, the metric exhibits a naked singularity.

\subsection{Instanton}

We can obtain an Euclidean solution
 from solution~\eqref{eq:met6} by  Wick rotations, $t \to i \tau $
and $Q \to i q$. The latter is necessary in order to have a real
 metric. So the Euclidean solution reads
\begin{equation}\label{eq:met6beta}
\begin{aligned}
\dd s^2_6 = & 
\dfrac{1}{1+\alpha/r^2} \left[ \left( 1 - \big( r_h / r \big)^2 \right) + \dfrac{q^2}{8\pi^4} \dfrac{1}{r^4} \right] \dd \tau^2
+ \dfrac{\dd r^2}{1 - \big( r_h / r \big)^2}
\\ &
+ r^2 \dd\Omega_3^2 
+ \Big(1+\alpha/r^2\Big) \dd y^2
- \dfrac{q}{\sqrt 2 \pi^2 r^2} \dd y \dd \tau \ .
\end{aligned}
\end{equation}
We stress that this is a solution of  pure (Euclidean) Einstein theory, 
where the action contains just the Einstein-Hilbert term,
\begin{equation}
I = \frac{1}{16 \pi G_6} \int\!\! \dd^6 x \, \left(\sqrt{g} R \right)_{\!6} \ ,
\end{equation}
without any additional field. For $q=0$, the solution reduces to the
Witten
instanton with an extra $S^1$ coordinate $y$. 

The geometry  has a potential conical singularity at the horizon which
must be removed as usual by a suitable identification.
The norm of the Killing vector, 
\begin{gather}
\zeta = \partial_\tau + A \, \partial_y \ , \\
A = \dfrac{q}{2\sqrt2 \pi^2} \dfrac{1}{\alpha + r_h^2} \ ,
\label{aaaa}
\end{gather}
vanishes at the horizon. 
We introduce a new coordinate $\tilde y = y - A \tau$, 
which is  constant along the orbits of $\zeta $. 
In terms of this coordinate, the metric takes the form
%
\begin{equation}
\begin{aligned}
\dd s^2_6 = &
\dfrac{1-\big(r_h/r\big)^2}{1+\alpha/r^2} \dd\tau^2 
+ \dfrac{\dd r^2}{1- \big(r_h/r\big)^2} + r^2 \dd\Omega_3^2 
\\ & 
+ \big( 1 + \alpha/r^2 \big) \left( \dd\tilde y
+ \frac{q}{2\sqrt2\pi^2} \left( \frac{1}{\alpha+r_h^2} 
- \frac{1}{\alpha+r^2}\right) \dd\tau \right)^{\!\!2} \ .
\end{aligned}
\end{equation}
Near the horizon the relevant part of the metric reduces to
\begin{equation}
\dd s^2 \approx \dfrac{2(r-r_h)}{r_h \big( 1 + \alpha/r_h^2 \big)} \dd\tau^2 
+ \dfrac{r_h}{2(r-r_h)} \dd r^2
+ \cdots \ .
\end{equation}
The horizon is at $r=r_h$. 
In order to avoid the conical singularity in~(\theequation), $\tau$
must be periodic, with $\tau = \tau + 2\pi R_\tau$ at {\it fixed}  $\tilde y$. 
The periodicity $2\pi R_\tau $ is determined by introducing a
Rindler coordinate $\tilde r=\sqrt{2 r_h}\sqrt{r-r_h}$.
The relevant part of the metric becomes 
\begin{equation}
\dd s^2 \approx \frac{\tilde r^2}{r_h^2+\alpha}\dd\tau^2+d\tilde r^2 + \cdots \ ,
\end{equation}
which shows that the space is free from conical singularity provided $\tau = \tau + 2\pi R_\tau$ with
\begin{equation}\label{eq:Rtaurh}
R_\tau =  \sqrt{  r_h^2+ \alpha  } \ .
\end{equation}
This gives the Hawking temperature $T_H=(2\pi R_\tau )^{-1}$.

At $r=\infty $, the instanton \eqref{eq:met6beta} approaches the metric
\beq
\dd s_6^2=\dd y^2+\dd \tau^2+\dd r^2+r^2 \dd\Omega_3^2\ ,
\eeq
with the identification 
\begin{equation}\label{eq:identif}
(y,\tau) = \Big( y + 2\pi n R_y + 2\pi m A R_\tau , \ \tau + 2\pi m R_\tau \Big) \ .
\end{equation}
Introducing $2\pi $-periodic variables $\sigma_1$, $\sigma_2$ by  $\tilde y=R_y\sigma_1$,
$\tau=R_\tau \sigma_2$, the asymptotic metric is
\beq
\dd s_6^2=R_y^2 \, \big|\dd \sigma_1+\Omega \dd \sigma_2 \big|^2  +\dd r^2+r^2 \dd\Omega_3^2\ ,
\label{torot}
\eeq
with $\Omega $ representing the torus modular parameter 
\beq
\Omega=a+i\, \frac{R_\tau}{R_y}\ ,\ \ \ \ a=\frac{AR_\tau}{ R_y}\ .
\label{modom}
\eeq
The behavior of the instanton at infinity tells us which is the
original vacuum that decays into the bubble. From~\eqref{torot} 
we see that the instanton approaches the space  $\mathbb R^{4} \times
T^2$, 
where the 2-torus has a modular parameter given by $\Omega$, eq.~\eqref{modom}.
The torus modular parameter is a feature 
which characterizes the Kaluza-Klein vacuum,
modulo $SL(2,\mathbb Z)$ transformations,
\beq
\Omega\to {a\Omega +b\over c\Omega+d}\ ,\ \ \ \ a,b,c,d\in \mathbb Z\
,
\ \ \ \ ad-bc=1\ .
\eeq
In particular, $a\to a+1$ gives an equivalent torus. This property is 
manifest in the identifications \eqref{eq:identif}, since 
$A\to A+R_y/R_\tau $ is absorbed into $n$.
We shall restrict the modular parameter $\Omega $ to the fundamental
domain ${\cal F}$ of $SL(2,\mathbb Z)$. In particular, this implies
the restriction
\beq
-{1\over 2}< a \leq {1\over 2}\ \ \ 
{\rm or}\ \ \  -{R_y\over 2R_\tau } < A \leq {R_y\over 2R_\tau }\ .
\label{restric}
\eeq


It is convenient to write the instanton metric~\eqref{eq:met6beta} in terms of the
parameters $R_\tau$, $R_y$,
$A$ that specify the Kaluza-Klein vacuum.
Using eqs.~\eqref{QK}, \eqref{aaaa}, \eqref{eq:Rtaurh} we find 
\begin{equation}
k = \dfrac{2 A^2}{3 + A^2} \ .
\end{equation}
Therefore
\begin{subequations}\label{eq:paramsbeta}
\begin{align}
q & = 2\sqrt{2}\pi^2 R_\tau^2 A\ , \\
r_h^2 & =  R_\tau^2 (1+A^2)\ , \\
\alpha & = -  R_\tau^2 A^2 \ .
\end{align}
\end{subequations}
%
%
The instanton metric becomes
\begin{equation}
\label{eq:reductytilde}
\begin{aligned}
\dd s^2_6 = &
\dfrac{1-(1+A^2) R_\tau^2/r^2}{1- A^2R_\tau^2/r^2} \dd\tau^2 
+ \dfrac{\dd r^2}{1-(1+A^2) R_\tau^2/r^2} + r^2 \dd\Omega_3^2 
\\ & 
+ \Big( 1 -  A^2R_\tau ^2/r ^2 \Big) \left( \dd\tilde y
+  A \, \dfrac{1-(1+A^2) R_\tau^2/r^2}{1- A^2R_\tau^2/r^2} \, \dd\tau \right)^{\!2} \ .
\end{aligned}
\end{equation} 
The metric is  regular on the full (geodesically complete)
space  $r^2\geq (1+A^2)R_\tau^2$.
Due to fact that the shift $A\to A+nR_y/R_\tau $ gives
an equivalent Kaluza-Klein vacuum at infinity, 
there will be an infinite family of instanton solutions
obtained by replacing $A\to A+nR_y/R_\tau $ in the metric 
\eqref{eq:reductytilde}.

Reducing  \eqref{eq:reductytilde} in  $\tilde y$, i.e. along the orbits of the $\zeta$ Killing vector, we find 
\begin{subequations}
\begin{align}
\phi 
& = -\frac12\ln\Big( 1 - (R_\tau A /r)^2 \Big) \ , \\
\tilde A_\tau 
& =  A \, \dfrac{1-(1+A^2) R_\tau^2/r^2}{1- A^2R_\tau^2/r^2} \ , \label{eq:Atilde}\\
\dd s^2_5 
& = \dfrac{1-(1+A^2)\big(R_\tau/r\big)^2}{\Big(1-\big(A R_\tau/r\big)^2\Big)^{2/3}} \dd\tau^2 
+ \dfrac{\Big(1-(A R_\tau/r)^2\Big)^{\!1/3}}{1- (1+A^2)\big(R_\tau/r\big)^2} \dd r^2
\nonumber\\&\quad+  \Big(1-(A R_\tau/r)^2\Big)^{\!1/3} r^2 \dd\Omega_3^2
\ . \label{eq:met5beta}
\end{align}
\end{subequations} 
From the point of view of the dimensionally 
reduced theory, the torus tilt is a non-zero gauge potential at infinity,
\begin{equation}
\tilde A_\tau(\infty) =A\ .
\end{equation}

\subsection{Bubble decay}
\def\t{{\rm t}}

In the standard treatment of semiclassical vacuum
 decay~\cite{Coleman:1977py,Coleman:1980aw},
the false vacuum decays into a Lorentzian space that coincides with
 the instanton (the bounce solution) on a three-dimensional surface
 of zero extrinsic curvature at $\t = 0$. Any Euclidean
 solution of finite action that has the same asymptotic as the Kaluza-Klein vacuum and
 that can be analytically continued to a real Lorentzian metric
 represents a decay mode of the original space.


The Lorentzian signature metric is obtained by a suitable analytic continuation of our instanton solution~\eqref{eq:met6beta}. The new time variable has to preserve the symmetry around the hyperspace of $\t =0$, so, 
as in \cite{Witten:1981gj}, the $\theta = \pi / 2$ plane can be taken  as the $\t =0$ surface. We write
\begin{equation}
\dd \Omega_3^2 = \dd\theta^2 + \sin^2\theta\, \dd\Omega_2^2 \ ,
\end{equation}
and perform a Wick rotation $\theta \to \pi/2 + i\psi$. 
This leads to the replacement
\begin{equation}
\dd\Omega^2_3 \to -\dd\psi^2 + \cosh^2\psi\, \dd\Omega^2_2 \ . 
\end{equation}
The resulting six dimensional metric contains the terms (see~\eqref{eq:reductytilde})
$$
\dd s_6^2=\frac{\dd r^2}{ 1- \big(1+A^2\big){R_\tau ^2}/{r^2}}-r^2\dd \psi^2+r^2
\cosh^2\psi \, \dd\Omega^2_2  +...
$$
Introducing coodinates
\beq
\rho  = r \cosh \psi \ , \ \ \ \ \t  = r \sinh \psi \ ,
\eeq
at large $r$ the metric approaches, 
\begin{equation}
\dd s^2_6 \cong  - \dd \t^2 + \dd\rho^2 +  \rho^2\dd\Omega_2^2 +\dd\tau^2 + \big( \dd\tilde y + A  \dd\tau \big)^2 \ .
\end{equation}
So the space at large $r$ is the usual flat space 
$\mathbb M^4 \times T^2$ with a
torus tilt. 
The full space at time $\t $ is a curved spacetime where the region
$\rho^2-\t ^2<r_h^2$, with $r_h^2=(1+A^2)R_\tau^2$, has been removed.
$r = r_h$ is a frontier of the space time: the wall of the bubble.
The space is regular at $r \ge r_h$.

The  radius of the bubble grows with time as
\begin{equation}
\rho_\text{bubble}(t) =  \sqrt{ \vphantom{r_h^2} (1+A^2)R_\tau^2 + \t^2 }\ .
\end{equation}
Note that the size of the bubble at $\t=0$ increases with the torus tilt parameter $A$.

The size of the $\tau$-circle at constant $\tilde y$ can be read directly from~\eqref{eq:reductytilde}
\begin{equation}
R^2_\tau(r)  = R_\tau^2  \, \big( 1 + A^2\big)\, 
\left(1-(1+A^2) {R_\tau^2\over r^2} \right)=
r_h^2 \, \left(1- {r_h^2 \over r^2}\right)\ .
\end{equation}
%
Thus the $\tau$-circle smoothly shrinks to zero at the surface of the
bubble. 
At infinity, it is $R_\tau(\infty )^2=R_\tau^2(1+A^2)$, as follows also
from \eqref{torot} by looking at the size of the $\sigma_2$ circle at
 $\sigma_1=$\~constant.

The size of $\tilde y $ circle is given by 
\begin{equation}
R_y(r) = R_y \, \sqrt{1-\big( A R_\tau/r \big)^{2}}\ ,
\end{equation}
with $r^2=\rho^2-\t^2>(1+A^2)R_\tau^2$, so that $R_y/\sqrt{1+A^2}\leq R_y(r)\leq R_y$.

%

\section{The instanton action}

In the semiclassical approximation,
the decay rate can be written
 as the exponential of minus the instanton action,
\begin{equation}
\Gamma \propto \ee{-I} \ .
\end{equation}
 We provide two independent
computations of the action. By direct evaluation
of the action including boundary terms, and by using thermodynamics.

\subsection{Direct calculation} 

In six dimensions, the only field we have is the metric, 
so the action is given by 
\begin{equation}\label{eq:action6}
I = -\frac{1}{16\pi G_6} \int \!\! \dd^6 x \left( \sqrt{g} R \right)_{\!6} - \frac1{8\pi G_6} \oint\! \sqrt{h} \big( K - K_0 \big) \ ,
\end{equation}
where $h$ is the induced boundary metric, $K$ the trace of its
extrinsic curvature and $K_0$ the analog quantity for the space where
our metric is embedded in (flat space, in our case). 
Since our instanton is a vacuum solution, $R_{\mu\nu} = 0$,  the only
contribution will come from 
the boundary term. 

If $n^\mu$ is a unit normal vector to the boundary, the extrinsic
curvature is written as 
%
\begin{equation}
K_{\mu\nu} = \nabla_\mu n_\nu - n_\mu n^\sigma \nabla_\sigma n_\nu \ .
\end{equation}
If the metric is such that $n^\mu$ is radial and only depends on the
radial coordinate, one can show that the trace of~(\theequation)
reduces to the logarithmic derivative of the square root of the
determinant of the boundary metric with respect to a unit radial
vector,
\begin{equation}
\sqrt{h} \, K = -\dfrac{1}{\sqrt{g_{rr}}} \partial_r \sqrt{h} \ .
\end{equation}
In our case, the metric induced at the boundary is given by~\eqref{eq:met6beta} with $r\to\infty$ and constant. We get
\begin{equation}
\frac{\partial}{\partial r} \sqrt{h} = \sin^2\theta_1 \sin\theta_2 \Big( 3 r^2- \frac12 r_h^2 \Big) + O(r^{-2}) \ .
\end{equation}
The flat space term, $K_0$, can be obtained from~(\theequation) imposing $q = M = 0$, i.e. $r_h = 0$.
 So we have
\begin{equation}
\sqrt h (K-K_0) = -\frac12 \sin^2\theta_1 \sin\theta_2 \, r_h^2 \ .
\end{equation}
Thus, the action is simply the integral of~(\theequation),
\begin{equation}\label{eq:Ibrute}
I = \frac1{8\pi G_6}  \pi^2 (2\pi R_y) (2\pi R_\tau) r_h^2 \ .
\end{equation}
Using $T = 1/(2\pi R_\tau)$ and~\eqref{eq:Rtaurh} we obtain
\begin{equation}
I = (2\pi R_y)\frac{\pi^2}{4 G_6} r_h^3 \sqrt{1+ \alpha/r_h^2}
=\frac{\pi^2 R_\tau^3}{4 G_5} (1+A^2) \ .
\end{equation}
where we have used $G_5 = G_6/(2\pi R_y)$.

\subsection{Computation using thermodynamics}

In five dimensions, the instanton metric represents an Euclidean
charged black hole, whose thermodynamics is well known.
The bulk part of the action~\eqref{eq:action6} becomes,
\begin{equation}
I = -\frac{1}{16\pi G_5} \int \!\! \dd^5 x  
\sqrt{-g_5} \left(R_5 -\frac{4}{3} (\partial \phi)^2 - \frac14 \ee{-8\phi/3} F_{\mu\nu}F^{\mu\nu} \right) \ .
\end{equation}
The action is the ratio between the free energy $F$ 
and the temperature $T=(2\pi R_\tau )^{-1}$. 
The free energy is
\begin{equation}
F = M - T S - \Phi_H Q \ ,
\end{equation}
where $\Phi_H$ is electric potential at $r = r_h$. We can simplify
this equation using the Smarr formula \cite{Gibbons:1987ps},
%
\begin{equation}
M = \frac{3}{2} \, T S + \Phi_H Q \ .
\end{equation}
Thus the action reads
\begin{equation}
I = \frac{F}{T} = \frac12 \, S \ .
\end{equation}
The entropy of the five dimensional black hole is the area of the
horizon over $4 G_5$, 
\begin{equation}
S = \dfrac{2\pi^2 }{4 G_5}\, r_h^3 \sqrt{1+\alpha/r_h^2 } 
= \dfrac{\pi^2 }{2 G_5}\, R_\tau^3 (1+A^2)
\ ,
\end{equation}
where we have used~\eqref{eq:met5beta}. So the action is
\begin{equation}
I = \frac{\pi^2 R_\tau^3}{4 G_5} (1+A^2) \ ,
\end{equation}
in exact agreement with~\eqref{eq:Ibrute}.

\subsection{Decay rate}

The action is
\begin{equation}
I(R_\tau, R_y, A ) = I_0 \, \Big( 1 + A ^2 \Big) \ , 
\ \ \ \ -{R_y\over 2R_\tau } < A < {R_y\over 2R_\tau }\ ,
\label{asto}
\end{equation}
where
\begin{equation}
I_0 = \frac{\pi^3 R_\tau^3R_y}{2 G_6}=\frac{\pi R_\tau^2}{8 G_4}\
 ,\quad G_4=G_6/\big((2\pi R_\tau)(2\pi R_y)\big)
 \ .
\end{equation}
The Witten decay \cite{Witten:1981gj} corresponds to $A =0$, so that $I_{\rm Witten}=I_0$.
In this case the instanton is obtained by analytic continuation from the $D=5$ Schwarzschild black hole.
Remarkably, the instanton action $I$ increases with the torus tilt. This indicates that the effect of
the tilt \eqref{eq:identif} is to render the Kaluza-Klein vacuum more stable.
Intuitively, the reason is that the size of the bubble at the
moment of creation $ \t=0$ is larger the larger is the tilt,
so the cost of producing the bubble is greater.

One can also write down the action for the shifted instantons,
 $A\to A-nR_y/R_\tau $,
\begin{equation}
I(R_\tau, R_y, A ) = I_0 \, \left( 1 + \Big(A-{nR_y\over R_\tau}\Big) ^{\! 2} \right) \ \ ,
\label{sasto}
\end{equation}
We see that in the fundamental domain 
$ -{R_y\over 2R_\tau } < A < {R_y\over 2R_\tau }$ the dominant
decay mode (i.e. the one with less action) is $n=0$. Other decay modes are
exponentially suppressed.

\section{Decay of Type 0 String Theory}

In the presence of fermions, the instanton decay studied here (just as in \cite{Witten:1981gj}) is relevant for
compactifications where
fermions obey antiperiodic boundary conditions in the $\tau $ direction.
The reason is that the space described by the instanton metric (topologically $\mathbb R^2\times S^1\times S^2$) 
admits a unique spin structure in the $\tau $-circle, since this shrinks to zero at the horizon.
There are two spin possible structures on $(\tau,\tilde y)$, namely $(-,+)$ and $(-,-)$, i.e.
 in the asymptotic region, fermions must be antiperiodic functions in the $S^1$ described by $\tau $.

As an application of the decay rate computed here, we consider
M-theory compactified on a two torus, where fermions obey antiperiodic
boundary conditions in one of the circles.
If we regard $\tau $ as the eleventh dimension, 
this compactification leads \cite{Gaber} to type 0A/0B theories \cite{SW}.
More precisely, ten-dimensional type 0B is equivalent to M-theory compactified on $T^2/[(-1)^F\times {\cal S}]$ in the limit of zero torus area, 
where $F$ is the spacetime fermion number and ${\cal S}$ is the half shift along the circle ($X\to X+\pi R$).
The type 0B Ramond-Ramond sector has an untwisted subsector $(R+,R+)$ and a twisted subsector $(R-,R-)$.
There are two RR scalars, one of each sector.
The tilt of the torus is related to the expectation value of the RR scalar of the untwisted sector.

For a rectangular torus, the RR scalar field vanishes.
The decay rate in this case was studied in \cite{costa}.
One assumes that there are six compact dimensions $y^m, \ m=1,...,6$
besides the 2-torus described by coordinates $\tau $ and $\tilde y$.
The relevant instantons describing the decay 
are essentially the Witten instanton  \cite{Witten:1981gj} and the
solution of  Dowker et al \cite{Dowker:1995}.
The relation with magnetic fields arises because
 string theory with antiperiodic fermions in one direction
can be described by a Melvin magnetic flux tube backgound with a special magnetic field \cite{RT}. For this critical value of the magnetic field, the dominant decay mode is  via bubble formation.
For small values of the magnetic field,
the decay is  dominated by creation of D6/D$\bar 6$-brane pairs, with a decay rate equal to the Schwinger rate 
\cite{costa} (see \cite{RT2,empgut} for related discussions). 
In addition to these
non-perturbative instabilities, type 0A theory at weak coupling $g_{0A}^2 =R_\tau^2/4\alpha'\ll 1$ has a tachyon
instability, which makes the theory highly unstable. In this regime,
the perturbative decay rate is of order one in $\alpha' $ units.

Now consider the decay in the presence of a torus tilt.
{}Consider first the case of  five extra directions $y^m$
compactified on a rectangular torus, so that we consider the decay
of the vacuum $\mathbb M^4\times T^2\times T^5$, and we wish to study
the effect of a tilt in $T^2$ on the decay rate.
The relevant instanton is the one constructed in the preceding sections,
by trivially adding five extra directions $y^m$. 
The action can be read from (\ref{asto}), 
%
\begin{equation}
\begin{aligned}
I_\text{0A/0B} &= \frac{\pi V_6R_\tau^2}{8 G_{10}} \, \Big( 1 + A^2\Big) =  \frac{4\pi V_6}{(2\pi)^6 {\alpha'}^3}
\left( 1 + a^2 \frac{R_y^2}{4\alpha' g_{0A}^2}\right)   \\ 
&=  \frac{4\pi V_6}{(2\pi)^6 {\alpha'}^3}
\left( 1 + \frac{a^2}{4 g_{0B}^2}\right)  \ , \ \ \ \ \ -{1\over
  2}<a\leq {1\over 2}\ ,
\end{aligned}
\end{equation}
%
where $G_{10}={G_{11}}\big/(2\pi R_\tau)$, $V_6=2\pi R_y V_5$ and $a$ is the expectation value of the vector component $A_{\tilde y} $ (or RR scalar in type 0B).
We have used
\begin{equation}
R_\tau^2=4\alpha' g_{0A}^2 \ ,
\quad 
16\pi G_{10}=(2\pi)^7 g_{0A}^2 {\alpha'}^4\ ,\quad g_{0B}=\frac{R_\tau}{ 2R_{ y}}\ ,
\quad a=\frac{A R_\tau }{ R_y}\ .
\end{equation}
Thus the decay rate  decreases as the tilt $a$ is increased.

Now consider the case where the five $y^m$ coordinates are non-compact.
The dominant non-perturbative decay
mode is again via the formation of a bubble of nothing.
For a generic torus with modular parameter $\Omega $, the relevant
instanton is obtained from the appendix by setting $D=10$
(this gives the instanton of maximal symmetry).  We thus obtain
the following decay rate of type 0A/0B theory:
\beq
\Gamma_{0A/0B}\propto e^{-I_{0A/0B}}\ ,\quad I_{0A/0B}=\frac{7^6\pi^4}{480 G_{10}}R_y R_\tau^7
\big(1+A ^2\big)^{7/2}\ ,  \ \ \ \ \ G_{10}=\frac{G_{11}}{ 2\pi R_\tau }\ .
\eeq
In terms of string theory parameters
\beq
I_\text{0A/0B}=\frac{7^6 g_{0A}^5 R_y}{30\pi^2\sqrt{\alpha' }} \left(1+  a^2 \frac{R_y^2}{4\alpha' g_{0A} ^2} \right)^{\!7/2} .
\label{swe}
\eeq
Interestingly, for a non-zero $a$ parameter  the
instanton action has a minimum, non-vanishing value 
as a function of the coupling $g_{0A}$,
 $I_{0A}^{\rm min}={\rm const.} R_y^6a^5/{\alpha'}^3$.
This means that for $ R_y^6a^5\gg {\alpha'}^3$, the action is large  and
the semiclassical approximation is reliable for any $g_{0A}$~.
At small couplings, there is 
a perturbative instability due to the presence of 
the type 0 tachyon which dominates over non-perturbative effects.
At strong coupling, $g_{0A}\gg 1$, the tachyon is expected to be absent, and the decay  should
be produced by the formation of the bubble of nothing, with a decay rate given by \eqref{swe}.
The decay is suppressed at large couplings. This is expected, since
at large couplings the dynamics of type 0A theory should approach that of
 M-theory, which is stable (for very large radius $R_\tau $, the
periodicity of the fermions should not significantly affect the
dynamics). 

The decay rate of type 0A theory at strong coupling when $\tilde y$ is 
non-compact 
is given by the Witten decay rate of a space $\mathbb M^{10} \times S^1$.
The rate can  be obtained from the formulas of the appendix
by  setting $D=11$, $A=0$ and replacing $G_{12}/2\pi R_y\to G_{11}$. We find
\beq
I_\text{0A}=\frac{\Omega_9}{ 8 G_{11}} 2^{16}R_\tau^9=\frac{2^{10}\pi^4 R_\tau^8}{ 3G_{10}}=\frac{2^{15}}{ 3\pi^2}\, g_{0A}^6\ .
\eeq
We see that the action increases with the string coupling to the power
six, so the decay rate is rapidly suppressed as the coupling is increased.




\acknowledgments
We would like to thank R. Emparan and L. Tagliacozzo for useful remarks.
This work is partially supported by CYT~FPA~2004-04582-C02-01 and 
CIRITGC 2001SGR-00065.
JLC is also supported by the Spanish's \textit{Ministerio de Educaci\'on y Ciencia}, FPU fellowship (ref:~AP2003-4193).

\bigskip

\appendix
\section{Decay of  $\mathbb M^{D-1} \times T^2$}
\label{app:Ddimensions}

\subsection{The $(D+1)$-dimensional instanton}

We can obtain the $(D+1)$-dimensional version of our instanton 
from the $D$-dimensional charged Kaluza-Klein black hole solution given in~\cite{Gibbons:1987ps}. 
By uplifting the solution to $D+1$ dimensions, and a Wick rotation along with $Q \to i q$, 
we find
\begin{equation}\label{eqappx:instantonD1}
\begin{aligned}
\dd s^2_\text{inst.}  = & \dfrac1{1+\alpha/r^{D-3}}
\left( \Big(1-\big(r_h/r)^{D-3}\Big) + \frac{2\, q^2 }{(D-3)^2 \Omega_{D-2}^2} \frac1{r^{2(D-3)}} \right) \dd \tau^2
\\ & 
+ \dfrac{\dd r^2}{ 1-\big( r_h/r \big)^{D-3} } 
+ r^2 \dd \Omega_{D-2}^2 
+ \Big( 1+\alpha/r^{D-3} \Big) \dd y^2 
\\ & 
- \frac{2 \sqrt{2} \, q}{(D-3) \Omega_{D-2}} \frac1{r^{D-3}} \dd y\,
\dd \tau \ ,\ \ \ \ \ \ \ \Omega_{D-2}=\frac{2\pi^{\frac{D-1} 2}}{\Gamma\big( \frac{D-1}{ 2} \big) }\ .
\end{aligned}
\end{equation}
Following the same procedure as in section~\ref{secc:solucion}, we
 consider
the  Killing vector $\zeta = \partial_\tau + A \partial_y$, where
\begin{equation}
A = \dfrac{\sqrt{2} q}{(D-3) \Omega_{D-2} r_h^{D-3}} 
\dfrac{1}{1 + \alpha \big/ r_h^{D-3}} \ , 
\end{equation}
which has zero norm at the horizon, and introduce the $\tilde y = y - A \tau$ variable, which is constant
 along
 the orbits of $\zeta$.
%
%
In order to remove the potential conical singularity, $\tau$ has to be
$(2\pi R_\tau)$-periodic at fixed $\tilde y$ coordinate, with $R_\tau$ 
now given by
\begin{equation}
R_\tau = \frac{2}{D-3}\, r_h \, \sqrt{1+\alpha/r_h^{D-3}} \ .
\end{equation}
Again, $A$ represents the value of the gauge field at infinity, $ \tilde A_\tau(\infty) = A$.
In order to express the solution in terms of  parameters $R_\tau$ and $A$,  it is first convenient to
 introduce
$k$ defined by 
\begin{equation}
k = 1 - \sqrt{1 - 2 \frac{D-2}{(D-3)^2} \left( \frac{q}{\kappa^2 M} \right)^{\!\!2} } \ .
\end{equation}
The  parameters are then given by
\begin{subequations}\label{eqappx:parsMKd}
\begin{align}
q & = (D-3) \, \kappa^2 M \sqrt{ \frac{k(2-k)}{2(D-2)}  } \ , \\
r_h^{D-3} & = \frac{2\kappa^2 M}{(D-2) \Omega_{D-2}} \left( 1 + \frac{D-3}{2} k \right)  \ , \qquad\qquad\qquad 
\alpha  = - \frac{\kappa^2 M}{\Omega_{D-2}} k \ , \\ 
\kappa^2 M & = \frac{(D-2) (D-3)^{D-3}}{2^{D-2}} \, \Omega_{D-2}  R_\tau^{D-3}
\left(1+ \frac{D-3}{2} k\right)^{\!(D-5)/2} \!
\left(1-\frac{k}{2} \right)^{\!\!-(D-3)/2}  \ .
\end{align}
\end{subequations}
The instanton solution is real for $k$ in the range 
 $0 \le k \le 2$ (so that $q$ is real).
Using 
\begin{equation}
A  = \sqrt{\frac{(D-2) k}{2-k}} \ ,
\end{equation}
i.e.
\begin{equation}
k = \dfrac{2 A ^2}{(D-2) + A ^2} \ .
\end{equation}
we obtain
\begin{subequations}\label{eqappx:parsA}
\begin{align}
q & = \frac{(D-3)^{D-2}}{2^{D-5/2}} \Omega_{D-2} R_\tau^{D-3} A \ \big(1+A^2\big)^{(D-5)/2} \ , \\
r_h & = \frac{D-3}{2} R_\tau \, \sqrt{1+A^2} \ ,\label{questa} \\
\alpha & = - \left( \frac{D-3}{2} R_\tau \right)^{D-3} A^2 \ \big( 1+A^2 \big)^{(D-5)/2} \ . 
\end{align}
\end{subequations}

At infinity, the instanton approaches the metric 
\beq
\dd s_{D+1}^2=\dd y^2+\dd \tau^2+\dd r^2+r^2 \dd\Omega_{D-2}^2\ ,
\eeq
with the identification 
\begin{equation}\label{identif}
(y,\tau) = \Big( y + 2\pi n R_y + 2\pi m A R_\tau , \ \tau + 
2\pi m R_\tau \Big) \ .
\end{equation}

\subsection{The instanton action}

In $D$ dimensions the solution represents an electrically charged black hole. The thermodynamics 
is as follows.
The Hawking temperature and entropy are
\begin{gather}\label{eqappx:temp}
T = \frac{1}{2\pi R_\tau} = \dfrac{D-3}{4\pi r_h \sqrt{1 + \alpha/r^{D-3}_h} } \ ,\\
S_D = \dfrac{\Omega_{D-2}\, r_h^{D-2}}{4G_D} \Big( 1+\alpha/r_h^{D-3} \Big)^{\!1/2} \ .
\end{gather} 
The action can be computed as the free energy over the temperature. 
Using 
$F = M - T S - \Phi_H Q $ and the Smarr formula
~\cite{Gibbons:1987ps} 
\begin{equation}
M = \frac{D-2}{D-3} T S + \Phi_H Q \ ,
\end{equation}
%
we find
\begin{equation}\label{eqappx:Itherm}
\begin{aligned}
I_\text{inst.}  & = \frac{F}{T}=\frac{S}{ D-3}   
= 2\pi R_y \, \frac{\Omega_{D-2}}{4G_{D+1}} \frac{r_h^{D-2}}{(D-3)}
\Big(1+\alpha/r^{D-3}_h\Big)^{\!1/2} \\
& = \frac{\Omega_{D-2}}{16\pi G_{D+1}} (2\pi R_y) (2\pi R_\tau) r_h^{D-3} 
\ ,
\end{aligned}
\end{equation}
where $G_{D+1}=2\pi R_y G_D$ and we have used \eqref{eqappx:temp}.

\medskip

The instanton action can also be computed directly from its definition as an integral,
\begin{equation}
I_\text{inst.} = - \frac{1}{16\pi G_{D+1}} \int\! \dd^{D+1}x  \, \sqrt{g} R - \frac{1}{8\pi G_{D+1}} \oint \! \sqrt{h} \big( K - K_0 \big) \ .
\end{equation}
Following the same method as in section 3.1, we obtain
%
%
\begin{equation}\label{eqappx:IKK0}
\begin{aligned}
I_\text{inst.} & =\frac1{16\pi G_{D+1}}\int\!\! \dd\Omega_{D-2} A(\theta_i) \int\!\! \dd \tau \dd y \, r_h^{D-3}\\
& = \frac{\Omega_{D-2}}{16\pi G_{D+1}} (2\pi R_y) (2\pi R_\tau) r_h^{D-3} \ .
\end{aligned}
\end{equation}
which exactly agrees with \eqref{eqappx:Itherm}.

\medskip

The decay rate is thus given by $\Gamma\sim e^{-I_\text{inst.}}$, with
\beq
I_\text{inst.}=I_0\ \big(1+A ^2\big)^{(D-3)/2} \ ,
\eeq
where
\beq
I_0=\frac{\pi\,\Omega_{D-2}}{ 4\ G_{D+1}} \left(\frac{D-3}{ 2}\right)^{\!\!D-3}\, R_y R_\tau^{D-2} \ .
\eeq
We have used \eqref{questa}.
The qualitative features are as in the decay of the six dimensional
space $\mathbb M^4 \times T^2 $.



%
%
%
\begingroup\raggedright
\endgroup
\end{document}